\documentclass[a4,11pt]{article}
\usepackage{graphicx} 
\usepackage{float}
\usepackage{amsmath}
\usepackage{amsfonts}
\usepackage{latexsym}
\usepackage{amssymb}
\makeatletter
 
  \@addtoreset{equation}{section}
 \makeatother

\begin{document}
\title{An FBSDE Approach to American Option Pricing\\
with an Interacting Particle Method~\footnote{
This research is supported by CARF (Center for Advanced Research in Finance) and 
the global COE program ``The research and training center for new development in mathematics.''
All the contents expressed in this research are solely those of the authors and do not represent any views or 
opinions of any institutions. 
The authors are not responsible or liable in any manner for any losses and/or damages caused by the use of any contents in this research.
}
%\\ \Large{\it{--Implications of asymmetric CSA and suboptimal strategies--}}
}

\author{Masaaki Fujii\footnote{Graduate School of Economics, The University of Tokyo},
Seisho Sato\footnote{The Institute of Statistical Mathematics}, 
Akihiko Takahashi\footnote{Graduate School of Economics, The University of Tokyo}
}
%\begin{center}
\date{
%First version: April 11, 2011\\
First version: November 26, 2012 %April 6 starts.
%Current version: \today
}
%\end{center}
\maketitle

%%%%%%    TEXT START    %%%%%%

%%%%%%      Macros      %%%%%%
%nakamacro.tex(H120522;0730)
%\documentstyle[11pt]{article}
%\setlength{\textwidth}{10.5in}
%\setlength{\oddsidemargin}{0in}
%\setlength{\topmargin}{-0.52in}
%\setlength{\textheight}{9.0in}
%\setlength{\footskip}{0.7in}

\newtheorem{definition}{Definition}
\newtheorem{assumption}{$[$ A}
\newtheorem{condition}{$[$ C}
\newtheorem{lemma}{Lemma}
\newtheorem{proposition}{Proposition}
\newtheorem{theorem}{Theorem}
\newtheorem{remark}{Remark}
\newtheorem{example}{Example}
\newtheorem{corollary}{Corollary}
%--------------------------------------------------------------------------
%BOLD FACES
\def\n{{\bf n}}
\def\A{{\bf A}}
\def\B{{\bf B}}
\def\C{{\bf C}}
\def\D{{\bf D}}
\def\E{{\bf E}}
\def\F{{\bf F}}
\def\G{{\bf G}}
\def\H{{\bf H}}
\def\I{{\bf I}}
\def\J{{\bf J}}
\def\K{{\bf K}}
\def\L{{\bf L}}
\def\M{{\bf M}}
\def\N{{\bf N}}
\def\O{{\bf O}}
\def\P{{\bf P}}
\def\Q{{\bf Q}}
\def\R{{\bf R}}
\def\S{{\bf S}}
\def\T{{\bf T}}
\def\U{{\bf U}}
\def\V{{\bf V}}
\def\W{{\bf W}}
\def\X{{\bf X}}
\def\Y{{\bf Y}}
\def\Z{{\bf Z}}
\def\cala{{\cal A}}
\def\calb{{\cal B}}
\def\calc{{\cal C}}
\def\cald{{\cal D}}
\def\cale{{\cal E}}
\def\calf{{\cal F}}
\def\calg{{\cal G}}
\def\calh{{\cal H}}
\def\cali{{\cal I}}
\def\calj{{\cal J}}
\def\calk{{\cal K}}
\def\call{{\cal L}}
\def\calm{{\cal M}}
\def\caln{{\cal N}}
\def\calo{{\cal O}}
\def\calp{{\cal P}}
\def\calq{{\cal Q}}
\def\calr{{\cal R}}
\def\cals{{\cal S}}
\def\calt{{\cal T}}
\def\calu{{\cal U}}
\def\calv{{\cal V}}
\def\calw{{\cal W}}
\def\calx{{\cal X}}
\def\caly{{\cal Y}}
\def\calz{{\cal Z}}
%
%YOKUTUKAUMONO
\def\sskip{\hspace{0.5cm}}
\def\simleq{ \raisebox{-.7ex}{\em $\stackrel{{\textstyle <}}{\sim}$} }
\def\leqsim{ \raisebox{-.7ex}{\em $\stackrel{{\textstyle <}}{\sim}$} }
\def\ep{\epsilon}
\def\half{\frac{1}{2}}
\def\iku{\rightarrow}
\def\Iku{\Rightarrow}
\def\ikup{\rightarrow^{p}}
\def\inclusion{\hookrightarrow}
\def\cadlag{c\`adl\`ag\ }
\def\up{\uparrow}
\def\down{\downarrow}
\def\doti{\Leftrightarrow}
\def\douti{\Leftrightarrow}
\def\dochi{\Leftrightarrow}
\def\douchi{\Leftrightarrow}%
%KAIGYOU,ARRAY
\def\yy{\\ && \nonumber \\}
\def\y{\vspace*{3mm}\\}
\def\nn{\nonumber}
\def\be{\begin{equation}}
\def\ee{\end{equation}}
\def\bea{\begin{eqnarray}}
\def\eea{\end{eqnarray}}
\def\beas{\begin{eqnarray*}}
\def\eeas{\end{eqnarray*}}
%
%KONO RONBUN DE TUKAU MONO
\def\hd{\hat{D}}
\def\hv{\hat{V}}
\def\hsd{{\hat{d}}}
\def\hx{\hat{X}}
\def\hsx{\hat{x}}
\def\bsx{\bar{x}}
\def\bsd{{\bar{d}}}
\def\bx{\bar{X}}
\def\ba{\bar{A}}
\def\bb{\bar{B}}
\def\bc{\bar{C}}
\def\bv{\bar{V}}
\def\balpha{\bar{\alpha}}
\def\bbalpha{\bar{\bar{\alpha}}}
\def\combi{\l(\begin{array}{c}\alpha\\ \beta \end{array}\r)}
\def\f{^{(1)}}
\def\s{^{(2)}}
\def\ss{^{(2)*}}
\def\l{\left}
\def\r{\right}
\def\a{\alpha}
\def\b{\beta}
\def\L{\Lambda}
%上に定義されたコマンドは数式モ−ドで用いる。
%--------------------------------------------------

\def\E{{\bf E}}
\def\P{{\bf P}}
\def\Q{{\bf Q}}
\def\R{{\bf R}}

\def\calf{{\cal F}}
\def\calp{{\cal P}}
\def\calq{{\cal Q}}

\def\ep{\epsilon}
\def\del{\delta}
\def\part{\partial}
\def\wh{\widehat}

\def\yy{\\ && \nonumber \\}
\def\y{\vspace*{3mm}\\}
\def\nn{\nonumber}
\def\be{\begin{equation}}
\def\ee{\end{equation}}
\def\bea{\begin{eqnarray}}
\def\eea{\end{eqnarray}}
\def\beas{\begin{eqnarray*}}
\def\eeas{\end{eqnarray*}}
\def\l{\left}
\def\r{\right}
\vspace{10mm}

%%%%%%%%%%%%%%%%%%%%%%%%%%%%%%$$$
\begin{abstract}
In the paper, we propose a new calculation scheme for American options in the framework of
a forward backward stochastic differential equation (FBSDE).
The well-known decomposition  of an American option price with that of a European option 
of the same maturity
and the remaining early exercise premium can be cast into the form of a decoupled non-linear FBSDE.
We numerically solve the FBSDE by applying an interacting particle method recently proposed by Fujii \& Takahashi (2012d),
which allows one to perform a Monte Carlo simulation in a fully forward-looking manner.
We perform the fourth-order analysis for the Black-Scholes (BS) model and the third-order analysis for the Heston model.
The comparison to those obtained from existing tree algorithms shows the effectiveness of the particle method.

\end{abstract}
\vspace{17mm}
%%%%%%%%%%%%%%%%%%%%%%%%%%%%%%%%%$
{\bf Keywords :}
BSDE, FBSDE, asymptotic expansion, perturbation, particle method
%%%%%%%%%%%%%%%%%%%%%%%%%%%%%%%%%

\newpage
%%%%%%%%%%%%%%%%%%%%%%%%%%%%%%%%%
\section{Introduction}
%%%%%%%%%%%%%%%%%%%%%%%%%%%%%%%%%
It has been known for a while that an American option value can be 
decomposed into that of the corresponding European option and 
an additional early exercise premium. 
Detailed discussions and other related references are available in Kim~(1990)~\cite{Kim},
Carr et.al.~(1992)~\cite{Carr}, Jacka~(1991)~\cite{Jacka}, Rutkowski~(1994)~\cite{Rutkowski},
Saito \& Takahashi~(2003)~\cite{Saito-T}
as well as a textbook written by Karatzas \& Shreve~(1998)~\cite{Karatzas}. 
See also a recent work of Benth~(2003)~\cite{Benth} which derives it from a 
dynamic programming approach.
In this paper, we deal with a non-linear forward backward stochastic differential equation (FBSDE)
obtained from the decomposition formula and calculate an American option price by solving it numerically.

The framework of FBSDE was first introduced by Bismut (1973)~\cite{Bismut}, 
and then later extended by Pardoux and Peng (1990)~\cite{P-Peng} for general non-linear cases.
Their financial applications are discussed in 
details in, for example, El~Karoui, Peng and Quenez (1997)~\cite{ElKaroui} and 
Ma and Yong (2000)~\cite{Ma}. 
There are increasing interests among researchers in FBSDEs since their 
relevance for the analysis of various social phenomena is becoming 
more apparent in recent years. In fact, one can find FBSDEs in the valuation problem of the financial contracts 
in the presence of credit risk and/or funding 
cost of collaterals~( Duffie \& Huang (1996)~\cite{Duffie}, Fujii \& Takahashi (2012a)~\cite{cva_asymmetric}, Cr\'epey (2012)~\cite{crepey}, for examples.~).  They are also relevant for the utility-indifference pricing in incomplete as well as
constrained markets ( Carmona (2009)~\cite{Carmona} and references therein.~). In a recent book of 
Cvitani\'c and Zhang (2012)~\cite{Cvitanic}, the authors use FBSDEs to study 
the optimal contract theory in continuous time. 

Recently, Fujii \& Takahashi (2012b)~\cite{FBSDE_approx} has proposed a  
perturbative technique for generic non-linear FBSDEs. 
With the help of asymptotic expansion ( Takahashi (1999)~\cite{T} ), it is possible to derive
closed-form analytic expressions for both of the backward components. An explicit example for 
a quadratic-growth FBSDE appearing in the optimal portfolio problem in an incomplete market 
is available in Fujii \& Takahashi (2012c)~\cite{qgFBSDE}.
In the following paper, Fujii \& Takahashi (2012d)~\cite{Particle} gave its numerical evaluation 
scheme based on an interacting particle method inspired by the work of 
McKean (1975)~\cite{McKean},
\footnote{It is closely related to the research with a long history
on the branching Markov process and a certain class of semi-linear PDEs.
For instance, see
Fujita(1966)~\cite{fujita}, 
Ikeda, Nagasawa \& Watanabe(1965,1966,1968)~\cite{inw1},~\cite{inw2}, 
~\cite{inw3}, Ikeda et.al.(1966,1967)~\cite{inw4} 
and Nagasawa \& Sirao (1969)~\cite{Nagasawa}.}
which enables one to perform
Monte Carlo simulation in a fully forward-looking manner.~\footnote{
A related but different approach was recently applied to evaluate CVA by Henry-Labord\`ere (2012)~\cite{Labordere}.}
The validity of its approximation is discussed recently by Takahashi \& Yamada (2012b)~\cite{yamada} 
although it is still restricted to a decoupled non-linear
%in a restrictive 
setup. 
In the current paper, we apply this methodology to evaluate a non-linear FBSDE relevant for an American option.
Although there remains a small error when the option is far in the money, we shall see the effectiveness 
of the particle method in overall region. The current work not only gives a simple calculation scheme for 
American options but also serves as a concrete example showing the usefulness of the particle method 
to analyze non-linear FBSDEs and corresponding non-linear partial differential equations.
%%%%%%%%%%%%%%%%%%%%%%%%%%%%%%%%%%%%%%%%%
\section{FBSDE formulation}
%%%%%%%%%%%%%%%%%%%%%%%%%%%%%%%%%%%%%%%%%
Let us take the probability space as $(\Omega,\calf,\mathbb{Q})$,
where $\mathbb{Q}$ is a risk-neutral probability measure.
We consider a generic process for the relevant stock price as
\bea
dS_t=(r_t-y_t)S_t dt+S_t\sigma_t\cdot dW_t,
\eea
where $W$ is a $d$-dimensional $\mathbb{Q}$-Brownian motion 
and $\calf$ is a natural filtration generated by $W$.
All the stochastic processes are assumed to be $\calf_t$-adapted.
Here, $r$ and $y$ are processes for a risk-free interest rate and 
a dividend yield, respectively. $\sigma\in \mathbb{R}^d$ is a 
volatility process.

It is well-known (e.g.~\cite{Kim, Carr, Jacka, Rutkowski,Karatzas})
that the price of an American option on 
$S$ with a strike $K$ and an expiry $T$ can be expressed as
\bea
V_t&=&\beta_t\mathbb{E}\left[\left.\beta_T^{-1}\Psi^+(S_T)\right|\calf_t\right]+\beta_t\mathbb{E}\left[\left. \int_t^T \beta_u^{-1}C_u\bold{1}_{\{V_u\leq \Psi^+(S_u)\}}du
\right|\calf_t\right],
\label{BSDE_pre}
\eea
where $\Psi^+(x)=\max(\Psi(x),0)$ denotes a payoff function, which is
\be
\Psi(x)=\begin{cases}
		x-K & \text{for a Call} \nn \\
		K-x & \text{for a Put} \nn
		\end{cases}~.
\ee
$C_t$ is a process denoting an instantaneous early exercise premium
\be
C_t=\begin{cases}
	y_t S_t-r_t K & \text{for a Call} \nn \\
	r_t K-y_t S_t & \text{for a Put}
	\end{cases}
\ee
and 
\bea
\beta_t=\exp\Bigl(\int_0^t r_s ds\Bigr)
\eea
is a standard money-market account.

In the remaining part of this section, we provide a simple heuristic derivation 
of Eq.~(\ref{BSDE_pre}) for completeness. 
Firstly, let us provide the decomposition principle of the Snell envelope for a 
continuous semimartingale. 
\begin{proposition} Rutokowski (1994)~\cite{Rutkowski}\\
Suppose $X$ is a continuous semimartingale with canonical decomposition
\be
X=X_0+M+V
\ee
where $X_0$ is a constant, $M$ is a continuous local martingale with $M_0=0$, and
$V$ denotes a continuous finite variation process with $V_0=0$,
whose decreasing component satisfies $dV_t^d=\nu_t dt$ for some adapted 
nonnegative process $\nu$.
We assume that
the condition
\bea
\mathbb{E}\Bigl[ \sup_{0\leq t \leq T} |X_t|\Bigr]<\infty
\eea
is satisfied.  
Let $\{\tau^*_t\}_{t\in[0,T]}$ be a family of $\{\calf_t\}$-stopping times satisfying
\bea
\mathbb{E}[X_{\tau_t^*}]=\rm{ess}\sup_{t\leq \tau\leq T} \mathbb{E}[X_\tau|\calf_t],
\quad \forall t\in[0,T]~.
\eea
Then the following equation holds:
\bea
\mathbb{E}[X_{\tau_t^*}|\calf_t]=\mathbb{E}[X_T|\calf_t]
-\mathbb{E}\left[\left.\int_{\tau_t^*}^T\bold{1}_{\{\tau_u^*=u\}}dV_u\right|\calf_t\right]~.
\eea
\label{prop1}
\end{proposition}
Proof: See Appendix of \cite{Rutkowski}.
\\

For concreteness, let us choose a Call option with strike $K$ as an example.
We consider the dynamics of the discounted payoff process. By applying It\^o formula, 
we obtain
\bea
d\bigl(\beta_t^{-1}(S_t-K)^{+}\bigr)&=&-\beta_t^{-1}r_t(S_t-K)^+dt+
\beta_t^{-1}\left\{\bold{1}_{\{S_t\geq K\}}dS_t+\frac{1}{2}\del(X_t-K)d\langle S\rangle_t\right\} \nn \\
&=&\beta_t^{-1}\bold{1}_{\{S_t\geq K\}}S_t\sigma_t\cdot dW_t\nn \\
&&+\beta_t^{-1}\bold{1}_{\{S_t\geq K\}}(r_tK-y_tS_t)dt+\frac{1}{2}\beta_t^{-1}\del(S_t-K)d
\langle S\rangle_t,
\eea
where $\del(\cdot)$ is a Dirac delta function. More precisely speaking, the term involves
the delta function is represented by the local time. For our 
intuitive derivation, however, the Dirac delta function is more useful to borrow a clear 
economic insight in a later stage.
Now, applying Proposition~\ref{prop1} gives
\bea
V_t&=&\rm{ess}\sup_{t\leq \tau\leq T}\beta_t \mathbb{E}\Bigl[\beta_\tau^{-1} (S_\tau-K)^+\Bigr|\calf_t\Bigr]\nn \\
&=&\beta_t\mathbb{E}\Bigl[\beta_T^{-1}(S_T-K)^+\Bigr|\calf_t\Bigr]\nn \\
&&+\beta_t\mathbb{E}\left[\left.
\int_{\tau_t^*}^T\bold{1}_{\{\tau_u^*=u\}}\left\{
\beta_u^{-1}\bold{1}_{\{S_u\geq K\}}(y_u S_u-r_u K)du-
\frac{1}{2}\beta_u^{-1}\del(S_u-K)d\langle S\rangle _u\right\}\right|\calf_t\right]\nn \\
&=&\beta_t\mathbb{E}\Bigl[\beta_T^{-1}(S_T-K)^+\Bigr|\calf_t\Bigr]
+\beta_t\mathbb{E}\left[\left.
\int_{\tau_t^*}^T\bold{1}_{\{\tau_u^*=u\}}
\beta_u^{-1}\bold{1}_{\{S_u\geq K\}}(y_u S_u-r_u K)du\right|\calf_t\right],\nn 
\eea
where, in the second equality,  the last term vanishes due to the fact that the stock should be in-the-money
region $(S_u>K)$ when the option is early exercised.
It is now economically clear to see that the above result can be rewritten as
\bea
V_t=\beta_t\mathbb{E}\Bigl[\beta_T^{-1}(S_T-K)^+\Bigr|\calf_t\Bigr]
+\beta_t\mathbb{E}\left[\left.
\int_{t}^T \bold{1}_{\{V_u\leq (S_u-K)^+\}}\beta_u^{-1}(y_u S_u-r_u K)du\right|\calf_t\right]~.\nn 
\eea
Note that $\bold{1}_{\{V_u\leq (S_u-K)^+\}}\bold{1}_{\{S_u>K\}}=\bold{1}_{\{V_u\leq (S_u-K)^+\}}$ since the option value should 
always be positive.
For more rigorous treatment, see the related 
proof in \cite{Rutkowski, Karatzas} as well as \cite{Benth}.
The case for a Put option can be shown similarly. $\blacksquare$

Now, from Eq.~(\ref{BSDE_pre}), one can see
\bea
\beta_t^{-1}V_t+\int_0^t \beta_u^{-1}C_u\bold{1}_{\{V_u\leq \Psi^+(S_u)\}}du
\eea
is a $\mathbb{Q}$-martingale. Thus, we can conclude 
that the price of an American option satisfies
\bea
\begin{cases}
& dV_t=r_t V_t dt-C_t\bold{1}_{\{V_t\leq \Psi(S_t)\}}dt+Z_t\cdot dW_t \\
& V_T=\Psi^+(S_T) \\
& dS_t=(r_t-y_t)S_tdt+S_t\sigma_t\cdot dW_t,\qquad S_0=s 
\end{cases}
\label{BSDE}
\eea
where $Z\in\mathbb{R}^d$ is an appropriate $\calf_t$-adapted process that should be
solved at the same time with $V$. It is a non-linear FBSDE with a decoupled 
dynamics of forward component, or the stock process $S$.
Here, we have replaced $\Psi^+$ by $\Psi$ in the 
indicator function since $V$ should be clearly positive.
In the next section, we carry out perturbative approximation procedures to solve the 
above FBSDE.
%%%%%%%%%%%%%%%%%%%%%%%%%%%%%%%%%%%%%%%%%%%%%%%%%%%%%%%%%%%%%%%%
\section{Perturbative expansion and a particle method for FBSDE}
%%%%%%%%%%%%%%%%%%%%%%%%%%%%%%%%%%%%%%%%%%%%%%%%%%%%%%%%%%%%%%%%
In \cite{FBSDE_approx}, a systematic 
approximation procedures for a generic non-linear FBSDE is given.
It treats the non-linear driver of the FBSDE as a perturbation and 
converted the original system into a series of decoupled linear FBSDEs,
for which the issue is equivalent to solve general European contingent claims.

To apply the procedures, let us introduce perturbation parameter $\ep$ as
\be
\begin{cases}
& dV_t^{(\ep)}=r_t V_t^{(\ep)}dt-\ep C_t\theta(\Psi(S_t)-V_t^{(\ep)})dt+Z_t^{(\ep)}\cdot dW_t \\
&V_T^{(\ep)}=\Psi^+(S_T)
\label{p_BSDE}
\end{cases}
\ee
where $\theta(\cdot)$ is the Heaviside step function. 
We now suppose that the solution of $(\ref{p_BSDE})$ can be expanded as a power series of $\ep$:
\bea
V_t^{(\ep)}&=&V_t^{(0)}+\ep V_t^{(1)}+\ep^2 V_t^{(2)}+\ep^3 V_t^{(3)}+\cdots \nn \\
Z_t^{(\ep)}&=&Z_t^{(0)}+\ep Z_t^{(1)}+\ep^2 Z_t^{(2)}+\ep^3 Z_t^{(3)}+\cdots ~.\nn 
\eea
Economically speaking, we treat the early exercise premium as a perturbation and 
expand the price of American option around the corresponding European price.
The method~\cite{FBSDE_approx} allows to derive a series of linear FBSDEs specifying 
the dynamics of $(V^{(i)},Z^{(i)})_{i\geq 0}$ for each order of $\ep$.
If the non-linear effects are sub-dominant and allow perturbative treatments,
we can expect to obtain a reasonable approximation of the original model
by setting $\ep=1$ at the end of the calculations.
For the evaluation of an American option, the driver (or drift term) of the FBSDE is independent of
the martingale component $Z$.  Thus, in the following, we can focus on the level component $V$. 

%%%%%%%%%%%%%%%%%%%%%%%%%%%%%%%%%%%%%%%%%%%%
\subsection{$0$th order}
%%%%%%%%%%%%%%%%%%%%%%%%%%%%%%%%%%%%%%%%%%%%
In the $0$th order, we have
\bea
\begin{cases}
& dV_t^{(0)}=r_t V_t^{(0)}dt+Z_t^{(0)}\cdot dW_t \\
& V_T^{(0)}=\Psi^+(S_T)
\end{cases}
\eea
which clearly represents the dynamics of the corresponding European option price.
We can easily see that it is solved as
\bea
V^{(0)}_t=\beta_t\mathbb{E}_t\left[\left. \beta_T^{-1}\Psi^+(S_T)\right|\calf_t\right]~.
\label{European}
\eea
Although there is no explicit expression of (\ref{European}) for a generic stock process,
it is always possible to obtain its approximation by asymptotic expansion (See \cite{T,KT,asymptotic3,asymptotic4}
for the details of asymptotic expansion.). It allows us, at least approximately, to have
an explicit expression of $V_t^{(0)}$ as
\be
V_t^{(0)}=v^{(0)}(t,\mathbf{X}_t)
\ee
where $\mathbf{X}_t=(S_t,r_t,y_t,\sigma_t,\cdots)$ contains all the relevant state processes.
If necessary, application of It\^o formula or using the process of Malliavin derivative ($\cald_t X_t$)
yields the corresponding martingale component $Z^{(0)}$.

%%%%%%%%%%%%%%%%%%%%%%%%%%%%%%%%%%%%%%%%%%%%%%%%
\subsection{$1$st order}
%%%%%%%%%%%%%%%%%%%%%%%%%%%%%%%%%%%%%%%%%%%%%%%%
In the $1$st order, the relevant FBSDE is given by
\bea
\begin{cases}
& dV_t^{(1)}=r_tV_t^{(1)}dt-C_t\theta\Bigl(\Psi(S_t)-v^{(0)}(t,\mathbf{X}_t)\Bigr)dt+Z_t^{(1)}\cdot dW_t \\
& V_T^{(1)} = 0
\end{cases}
\eea
which is again linear and easy to integrate.
We have
\bea
V_t^{(1)}=\int_t^T du \beta_t \mathbb{E}\Bigl[\beta_u^{-1}C_u\theta(\Psi(S_u)-v_u^{(0)})\Bigr|\calf_t\Bigr]~
\label{V1exp}
\eea 
where, $v_u^{(0)}$ denotes $v^{(0)}(u,\mathbf{X}_u)$.
$Z^{(1)}$ is obtained by the similar arguments given in the previous subsection.
Although it is possible to evaluate (\ref{V1exp}) directly by Monte Carlo simulation,
the time integration makes it rather time consuming. In fact, it soon becomes infeasible 
when one evaluates $\ep$-higher order expansion terms.

In order to avoid the difficulty, we adopt an interacting particle method proposed in 
Fujii \& Takahashi (2012d)~\cite{Particle}. We introduce an arbitrary $\calf_t$-adapted
strictly positive process $\{\lambda_t\}_{t\geq 0}$ to define  
\be
\wh{V}_{t,s}^{(1)}=\exp\Bigl(\int_t^s \lambda_u du\Bigr)V_s^{(1)}
\ee
and 
\be
\wh{C}_{t,s}=\frac{1}{\lambda_s}\exp\Bigl(\int_t^s \lambda_u du\Bigr)C_s
\ee
for $s\geq t$.
Then, we have the SDE of $\wh{V}^{(1)}_{t,s}$ for the time component $s~(\geq t)$,
\bea
\begin{cases}
&d\wh{V}_{t,s}^{(1)}=(r_s+\lambda_s) \wh{V}_{t,s}^{(1)}ds-\lambda_s\wh{C}_{t,s}\theta(\Psi(S_s)-v_s^{(0)})ds+
e^{\int_t^s \lambda_u du}Z_s^{(1)}\cdot dW_s \\
&\wh{V}^{(1)}_{t,T}=0 
\end{cases}.
\eea
Since $\hat{V}_{t,t}^{(1)}=V_t^{(1)}$, we have
\bea
V_t^{(1)}&=&\mathbb{E}\left[\left.\int_t^T e^{-\int_t^s (r_u+\lambda_u)du}\lambda_s \wh{C}_{t,s}\theta(
\Psi(S_s)-v_s^{(0)})ds\right|\calf_t\right] \\
&=&\bold{1}_{\{\tau_1>t\}}\mathbb{E}\left[\left.
\bold{1}_{\{\tau_1< T\}}e^{-\int_t^{\tau_1}r_u du}\wh{C}_{t,\tau_1}\theta(\Psi(S_{\tau_1})-v_{\tau_1}^{(0)})
\right|\calf_t\right]~. 
\label{V1st}
\eea
Here,  $\tau_1$ is a $\calf_t$-stopping time associated with the first jump of Poisson process 
whose intensity process is given by $\{\lambda_t\}_{t\geq 0}$.
In contrast to (\ref{V1exp}), it is clear the expression of (\ref{V1st}) allows one-shot Monte Carlo simulation.
More detailed explanation for Monte Carlo simulation will be given in the later section.
Although it is an interesting topic to obtain an optimal intensity process $\lambda$
that achieves the smallest variance in simulation, it is beyond the current scope of the paper.
In the numerical examples, we simply use a constant intensity.
\\
\\
{\bf{Remark}:}~In \cite{Particle}, the intensity process $\lambda$ is assumed to be deterministic or
an independent process for the other underlyings, which makes the evaluation of Malliavin derivatives required for $Z^{(i)}$
simpler. For the evaluation of American option, this assumption is not necessary since there is no need to obtain
$Z^{(i)}$.

%%%%%%%%%%%%%%%%%%%%%%%%%%%%%%%%%%%
\subsection{$2$nd order}
%%%%%%%%%%%%%%%%%%%%%%%%%%%%%%%%%%%
For the 2nd order case, the relevant equation is given by
\bea
\begin{cases}
&dV_t^{(2)}=r_tV_t^{(2)}dt+C_t\del\Bigr(\Psi(S_t)-v_t^{(0)}\Bigr)V_t^{(1)}dt+Z_t^{(2)}\cdot dW_t \\
&V_T^{(2)}=0
\end{cases}
\eea
where $\del(\cdot)$ is a Dirac delta function as before.
Since the FBSDE is linear, one can show easily that
\bea
V_t^{(2)}=-\beta_t \int_t^T du  \mathbb{E}\left[\left.\beta_u^{-1}C_u\del\Bigl( \Psi(S_u)-v_u^{(0)}
\Bigr)V_u^{(1)}\right|\calf_t\right]~.
\label{V2ndDirect}
\eea
As mentioned in the previous section, the difficulty in a naive application of Monte Carlo simulation becomes 
much clearer now. At each point of time $u\in[t,T]$ in a given path, one needs the value of $V_u^{(1)}$, 
which in turn requires to run Monte Carlo simulation as well as time integration. 

Therefore, let us define
\bea
&&\wh{V}_{t,s}^{(2)}=\exp\Bigl(\int_t^s \lambda_u du\Bigr) V_s^{(2)} 
\eea
and use $\wh{C}_{t,s}$ as before. Then, for $s\geq t$, we have
\bea
d\wh{V}_{t,s}^{(2)}=(r_s+\lambda_s)\wh{V}_{t,s}^{(2)}ds+\lambda_s\wh{C}_{t,s}\del\Bigl(\Phi(S_s)-v_s^{(0)}\Bigr)V_s^{(1)}ds
+e^{\int_t^s\lambda_u du}Z_s^{(2)}\cdot dW_s 
\eea
with $\wh{V}_{t,T}=0$. Thus, one obtains
\bea
&&V_t^{(2)}=\wh{V}_{t,t}^{(2)}\nn \\
&&=-\mathbb{E}\left[\left.\int_t^T e^{-\int_t^s(r_u+\lambda_u)du}\lambda_s\wh{C}_{t,s}\del
(\Psi(S_s)-v_s^{(0)})V_s^{(1)}ds\right|\calf_t\right]\nn \\
&&=-\bold{1}_{\{\tau_1>t\}}\mathbb{E}\left[\left.\bold{1}_{\{\tau_1<T\}}
e^{-\int_t^{\tau_1}r_u du}\wh{C}_{t,\tau_1}\del(\tau_1)V_{\tau_1}^{(1)}\right|\calf_t\right]~.
\eea
Simple application of the tower property of iterated expectations gives
\bea
V_t^{(2)}=-\bold{1}_{\{\tau_1>t\}}\mathbb{E}\left[\left.
\bold{1}_{\{\tau_1<\tau_2<T\}}e^{-\int_t^{\tau_2}r_u du}\wh{C}_{t,\tau_1}\del(\tau_1)
\wh{C}_{\tau_1,\tau_2}\theta(\tau_2)\right|\calf_t\right]~,
\label{V2nd}
\eea
where $\tau_1$ ($\tau_2$) is the first (second) jump time of the Poisson process with the 
intensity process $\lambda$.
Here, we have written $\del(\tau)=\del(\Psi(\tau)-v_{\tau}^{(0)})$ and 
$\theta(\tau)=\theta(\Psi(\tau)-v_{\tau}^{(0)})$ to lighten the notations.

%%%%%%%%%%%%%%%%%%%%%%%%%%%%%%%%%%%%%%%%%%%
\subsection{$3$rd order}
%%%%%%%%%%%%%%%%%%%%%%%%%%%%%%%%%%%%%%%%%%%
In the 3rd order, the relevant dynamics becomes
\bea
\begin{cases}
&dV_t^{(3)}=r_tV_t^{(3)}dt+C_t\Bigl\{ \del\Bigl(\Psi(S_t)-v_t^{(0)}\Bigr)V_t^{(2)}\nn \\
&\hspace{30mm}-\frac{1}{2}\part \del\Bigl(\Psi(S_t)-v_t^{(0)}\Bigr)(V_t^{(1)})^2
\Bigr\}dt+Z_t^{(3)} \cdot dW_t, \\
&V_T^{(3)}=0.
\end{cases}~
\label{V3bsde}
\eea
Here, the derivative of a Dirac delta function can be evaluated by approximating 
the delta function as a normal density function with a small variance, or using 
the integration-by-parts formula if possible. 
For the details of calculation, see the later sections treating numerical examples.
After integration, we obtain
\bea
V_t^{(3)}&=&-\beta_t \int_t^T du \mathbb{E}\Bigl[\beta_u^{-1}C_u\del(\Psi(S_u)-V_u^{(0)})V_u^{(2)}\Bigr|\calf_t\Bigr]\nn \\
&&+\beta_t\int_t^T du \mathbb{E}\Bigl[\beta_u^{-1}C_u\frac{1}{2}\part\del(\Psi(S_u)-V_u^{(0)})(V_u^{(1)})^2\Bigr|\calf_t\Bigr]~.
\eea

Let us compress a convoluted expectation as before.
Let us denote
\bea
\wh{V}_{t,s}^{(3)}=\exp\Bigl(\int_t^s \lambda_u du\Bigr)V_s^{(3)} 
\eea
and continue to use the simplified notations:
\bea
&&\theta(t)=\theta(\Psi(S_t)-v_t^{(0)})\\
&&\del(t)=\del(\Psi(S_t)-v_t^{(0)}) ~.
\eea
Then, (\ref{V3bsde}) is equivalent to
\bea
d\wh{V}_{t,s}^{(3)}=(r_s+\lambda_s)\wh{V}_{t,s}^{(3)}ds+\lambda_s\wh{C}_{t,s}\Bigl\{
\del(s)V_s^{(2)}-\frac{1}{2}\part\del(s)[V_s^{(1)}]^2\Bigr\}ds+e^{\int_t^s \lambda_u du}Z_s^{(3)}\cdot dW_s\nn
\eea
with $\wh{V}_{t,T}=0$, thus
\bea
&&V_t^{(3)}=\mathbb{E}\left[\left.\int_t^T e^{-\int_t^s (r_u+\lambda_u )du}
\lambda_s\wh{C}_{t,s}\Bigl\{-\del(s)V_s^{(2)}+\frac{1}{2}\part\del(s)[V_s^{(1)}]^2\Bigr\}ds\right|\calf_t\right]\nn \\
&&=\bold{1}_{\{\tau_1>t\}}\mathbb{E}\left[\left.\bold{1}_{\{\tau_1<T\}}
e^{-\int_t^{\tau_1}r_udu}\wh{C}_{t,\tau_1}\Bigl\{
-\del(\tau_1)V_{\tau_1}^{(2)}+\frac{1}{2}\part\del(\tau_1)[V_{\tau_1}^{(1)}]^2\Bigr\}\right|\calf_t\right]~.\nn \\
\eea
Borrowing the idea from McKean~\cite{McKean} and use the tower property of iterated expectations, we finally obtain
\bea
&&V_t^{(3)}=\bold{1}_{\{\tau_1>t\}}\mathbb{E}\left[\left. \bold{1}_{\{\tau_1<\tau_2<\tau_3<T\}}
e^{-\int_t^{\tau_3}r_udu}\wh{C}_{t,\tau_1}\del(\tau_1)\wh{C}_{\tau_1,\tau_2}\del(\tau_2)
\wh{C}_{\tau_2,\tau_3}\theta(\tau_3)\right|\calf_t\right]\nn\\
&&+\bold{1}_{\{\tau_1>t\}}\mathbb{E}\left[\left.\bold{1}_{\{\tau_1<T\}}\frac{1}{2}e^{-\int_t^{\tau_1}r_u du}
\wh{C}_{t,\tau_1}\part\del(\tau_1) \prod_{p=1}^2\left\{\bold{1}_{\{\tau_1<\tau_2^{(p)}<T\}}
e^{-\int_{\tau_1}^{\tau_2^{(p)}}r_u du}\wh{C}_{\tau_1,\tau_2^{(p)}}\theta(\tau_2^{(p)})\right\}\right|\calf_t\right]~\nn\\
\label{V3rd}
\eea
with the $i$-th jump time of the Poisson process denoted by $\tau_i$.

In (\ref{V3rd}), $p=\{1,2\}$ indicates one of the two particle groups. In both of the groups, the relevant state variables
(or particles) follow the common diffusion dynamics ( those specified by BS or Heston models, for example) 
but those belong to 
different groups are independent each other, i.e., driven by the two independent set of Brownian motions.
This particle representation compresses $(\mathbb{E}[~\cdot~|\calf_{\tau_1}])^2$
into a single expectation operation.

More concretely, for the evaluation of the second line, we use the branching diffusion method of McKean. 
For a each path of simulation, we\\
(1): update the diffusion process of the underlyings $X=\{S,r,y,\sigma,\cdots\}$ in a standard way. \\
(2): do Poisson draw with intensity $\lambda$ at each step.\\ 
(3): if it draws a "jump" (or particles interact) at $\tau_1<T$, then the path yields the two identical copies of 
particles $\{\mathbf{X}_p\}_{p=1,2}$ of the underlying states as its offspring, 
which continue to evolve according to the identical diffusion equations but driven by the two
independent set of Brownian motions. \\
(4): for each particle group, we continue the Poisson draw of the second interaction until the maturity.\\
(5): finally, extract the following term:
\bea
\bold{1}_{\{\tau_1<T\}}\frac{1}{2}e^{-\int_t^{\tau_1}r_u du}\wh{C}_{t,\tau_1}\part\del(\tau_1)
\prod_{p=1}^2\bold{1}_{\{\tau_1<\tau_2^p<T\}}e^{-\int_{\tau_1}^{\tau_2^p}r_u du}\wh{C}_{\tau_1,\tau_2^p}
\theta(\tau_2^p)
\eea
where $\tau_2^p$ is the second interaction time of each particle group.\\
(6): Repeat the procedures (1-5) and take the average of the values gathered in (5).

%%%%%%%%%%%%%%%%%%%%%%%%%%%%%%%%%%%%%%%%%%%%%%%%%%%
\subsection{$4$th order}
%%%%%%%%%%%%%%%%%%%%%%%%%%%%%%%%%%%%%%%%%%%%%%%%%%%
We can continue the expansion to an arbitrary higher order.
In the 4th order, we have
\bea
\begin{cases}
&dV_t^{(4)}=r_tV_t^{(4)}dt +C_t\left\{\frac{1}{3!}\part^2\del(\Psi(S_t)-v_t^{(0)})\bigl[V_t^{(1)}\bigr]^3-\part \del(\Psi(S_t)-v_t^{(0)})\bigl[V_t^{(1)}\bigr]\bigl[V_t^{(2)}\bigr] \right.\nn \\
&\hspace{40mm}\left.+\del(\Psi(S_t)-v_t^{(0)})\bigl[V_t^{(3)}\bigr]\right\}dt+Z_t^{(4)}\cdot dW_t \\
&V_T^{(4)}=0
\end{cases}
\eea
and hence
\bea
V_t^{(4)}&=&-\beta_t\int_t^T\mathbb{E}\left[\left.
\beta_u^{-1}\frac{1}{3!}C_u\part^2\del(\Psi(S_u)-v_u^{(0)})\bigl[V_u^{(1)}\bigr]^3\right|\calf_t\right]du\nn \\
&&+\beta_t\int_t^T \mathbb{E}\left[\left. \beta_u^{-1}C_u
\part \del(\Psi(S_u)-v_u^{(0)})\bigl[V_u^{(1)}\bigr]\bigl[V_u^{(2)}\bigr]\right|\calf_t\right]du \nn \\
&&-\beta_t\int_t^T \mathbb{E}\left[\left. \beta_u^{-1}C_u
\del(\Psi(S_u)-v_u^{(0)})V_u^{(3)}\right|\calf_t\right]du
\eea

Using the similar notations as in the previous sections, one can show that
\bea
&&d\wh{V}_{t,s}^{(4)}=(r_s+\lambda_s)\hat{V}_{t,s}^{(4)}ds+
\lambda_s\wh{C}_{t,s}\left\{\frac{1}{3!}\part^2\del(s)[V_s^{(1)}]^3-\part\del(s)[V_s^{(1)}][V_s^{(2)}]+\del(s)V_s^{(3)}
\right\}ds\nn \\
&&\qquad\qquad +e^{\int_t^s \lambda_u du}Z_s^{(4)}\cdot dW_s
\eea
and hence
\bea
&&V_t^{(4)}=\mathbb{E}\left[\left.\int_t^T e^{-\int_t^s(r_u+\lambda_u)du}\lambda_s\wh{C}_{t,s}
\left\{-\frac{1}{3!}\part^2\del(s)[V_s^{(1)}]^3+\part\del(s)[V_s^{(1)}][V_s^{(2)}]-\del(s)[V_s^{(3)}]\right\}ds
\right|\calf_t\right]\nn \\
&&=\bold{1}_{\{\tau_1>t\}}\mathbb{E}\left[\left.
\bold{1}_{\{\tau_1<T\}}e^{-\int_t^{\tau_1}r_s ds}\wh{C}_{t,\tau_1}\left\{
-\frac{1}{3!}\part^2\del(\tau_1)[V_{\tau_1}^{(1)}]^3+\part\del(\tau_1)[V_{\tau_1}^{(1)}][V_{\tau_2}^{(2)}]
-\del(\tau_1)[V_{\tau_1}^{(3)}]\right\}\right|\calf_t\right]\nn
\eea

Using the tower property and particle representation, the above result can be expanded as
\bea
&&V_t^{(4)}=-\bold{1}_{\{\tau_1>t\}}\mathbb{E}\left[\bold{1}_{\{\tau_1<T\}}\frac{1}{3!}e^{-\int_t^{\tau_1}r_sds}
\wh{C}_{t,\tau_1}\part^2\del(\tau_1)\prod_{p=1}^3
\left\{\bold{1}_{\{\tau_1<\tau_2^{(p)}<T\}}e^{-\int_{\tau_1}^{\tau_2^{(p)}}r_sds}
\wh{C}_{\tau_1,\tau_2^{(p)}}\theta(\tau_2^{(p)})\right\}\right.\nn \\
&&\qquad+\bold{1}_{\{\tau_1<T\}}e^{-\int_t^{\tau_1}r_sds}\wh{C}_{t,\tau_1}\part\del(\tau_1)
\left\{\bold{1}_{\{\tau_1<\tau_2^{(p)}<T\}}e^{-\int_{\tau_1}^{\tau_2^{(p)}}r_sds}\wh{C}_{\tau_1,\tau_2^{(p)}}
\theta(\tau_2^{(p)})\right\}^{p=1}\nn \\
&&\qquad\times\left\{\bold{1}_{\{\tau_1<\tau_2^{(p)}<\tau_3^{(p)}<T\}}
e^{-\int_{\tau_1}^{\tau_3^{(p)}}r_sds}\wh{C}_{\tau_1,\tau_2^{(p)}}\del(\tau_2^{(p)})\wh{C}_{\tau_2^{(p)},\tau_3^{(p)}}
\theta(\tau_3^{(p)})\right\}^{p=2}\nn \\
&&\qquad+\bold{1}_{\{\tau_1<\tau_2<\tau_3<\tau_4<T\}}e^{-\int_t^{\tau_4}r_sds}\wh{C}_{t,\tau_1}\del(\tau_1)
\wh{C}_{\tau_1,\tau_2}\del(\tau_2)\hat{C}_{\tau_2,\tau_3}\del(\tau_3)\hat{C}_{\tau_3,\tau_4}\theta(\tau_4)\nn \\
&&\qquad+\bold{1}_{\{\tau_1<\tau_2<T\}}\frac{1}{2}e^{-\int_t^{\tau_2}r_sds}
\wh{C}_{t,\tau_1}\del(\tau_1)\wh{C}_{\tau_1,\tau_2}\part\del(\tau_2)\nn \\
&&\qquad\times \prod_{p=1}^2\left.\left.\left\{
\bold{1}_{\{\tau_2<\tau_3^{(p)}<T\}}e^{-\int_{\tau_2}^{\tau_3^{(p)}}r_sds}\wh{C}_{\tau_2,\tau_3^{(p)}}\theta(\tau_3^{(p)})
\right\}\right|\calf_t\right]~.\nn\\
\label{V4th}
\eea

%%%%%%%%%%%%%%%%%%%%%%%%%%%%%%%%%%%%
\section{Numerical Examples}
%%%%%%%%%%%%%%%%%%%%%%%%%%%%%%%%%%%%%
This section demonstrates the validity of our method proposed in the previous section
through numerical experiments.
%%%%%%%%%%%%%%%%%%%%%%%%%%%%%%%%%%%%%%%%%%%%
\subsection{Example 1: Black-Scholes model}
%%%%%%%%%%%%%%%%%%%%%%%%%%%%%%%%%%%%%%%%%%%%
The first example is taken from Black-Scholes model:
\be
dS_t/S_t=(r-y)dt+\sigma dW_t,
\label{BSmodel}
\ee
where $r,y$ and $\sigma$ are all nonnegative constants.
We calculate the values up to the fourth order terms based on our scheme derived as (\ref{V1st}),
(\ref{V2nd}), (\ref{V3rd}) and (\ref{V4th}) with 10 million trials in Monte Carlo 
simulation.
Here, we adopt the values reported in ~\cite{Ju} as benchmarks.
In particular, difficulty arises 
in differentiations up to the second order of the delta functions required 
for evaluation of the third (\ref{V3rd}) as well as the fourth (\ref{V4th}) order terms.
Since the density function in Black-Scholes model is explicitly known,
we are able to apply integration by parts (IBP)
for computation of these terms
in order to avoid differentiation of the delta functions.
Moreover, we approximate each delta function by a normal density function
with mean zero and a certain variance, which enables direct evaluation of the expectation.

\begin{table}[H]
\footnotesize
\begin{center}
%\vspace{.3cm}\\																			
\caption{American Puts ($T=3, K=100, \sigma=0.2, r=0.08$) }\label{put3y}
\begin{tabular}{l|c|c|ccccc}																			
\hline																			
&	$S_0$	&	Benchmark	&	0th	&	1st	&	2nd	&	3rd	&	4th	\\
\hline
$y=0.12$&	80	&	25.658	&	24.777	&	25.829	&	25.854	&	25.799	&	25.739	\\
&	90	&	20.083	&	19.620	&	20.174	&	20.187	&	20.158	&	20.131	\\
&	100	&	15.498	&	15.252	&	15.546	&	15.553	&	15.538	&	15.518	\\
&	110	&	11.803	&	11.671	&	11.830	&	11.834	&	11.826	&	11.822	\\
&	120	&	8.886	&	8.814	&	8.900	&	8.902	&	8.897	&	8.894	\\
\hline
$y=0.08$&	80	&	22.205	&	19.525	&	23.553	&	22.847	&	22.265	&	22.194	\\
&	90	&	16.207	&	14.676	&	16.982	&	16.522	&	16.372	&	16.316	\\
&	100	&	11.704	&	10.817	&	12.151	&	11.885	&	11.800	&	11.735	\\
&	110	&	8.367	&	7.847	&	8.629	&	8.473	&	8.424	&	8.409	\\
&	120	&	5.930	&	5.622	&	6.081	&	5.989	&	5.962	&	5.951	\\
\hline
$y=0.04$&	80	&	20.350	&	14.589	&	23.683	&	22.236	&	21.450	&	20.348	\\
&	90	&	13.497	&	10.326	&	16.120	&	13.774	&	13.573	&	13.825	\\
&	100	&	8.944	&	7.168	&	10.390	&	9.132	&	9.070	&	8.706	\\
&	110	&	5.912	&	4.902	&	6.720	&	5.992	&	5.957	&	5.882	\\
&	120	&	3.898	&	3.315	&	4.360	&	3.953	&	3.928	&	3.827	\\
\hline
$y=0.00$&	80	&	20.000	&	10.253	&	24.338	&	22.044	&	20.892	&	20.063	\\
&	90	&	11.697	&	6.783	&	16.534	&	12.950	&	11.525	&	11.959	\\
&	100	&	6.932	&	4.406	&	9.590	&	6.719	&	7.004	&	6.697	\\
&	110	&	4.155	&	2.826	&	5.529	&	4.066	&	4.198	&	4.286	\\
&	120	&	2.510	&	1.797	&	3.232	&	2.457	&	2.506	&	2.582	\\
\hline
\end{tabular}\\
Number of simulations = 10,000,000, $\lambda=2$, Number of time steps = 6000
\vspace{15mm} ~ \\
Error ratio \\
\begin{tabular}{l|c|c|ccccc}														
\hline														
&	$S_0$	&	Benchmark	&	0th	&	1st	&	2nd	&	3rd	&	4th	\\
\hline														
$y=0.12$&	80	&	25.658	&	-3.434\%	&	0.666\%	&	0.764\%	&	0.550\%	&	0.316\%	\\
&	90	&	20.083	&	-2.305\%	&	0.453\%	&	0.518\%	&	0.373\%	&	0.239\%	\\
&	100	&	15.498	&	-1.587\%	&	0.310\%	&	0.355\%	&	0.258\%	&	0.129\%	\\
&	110	&	11.803	&	-1.118\%	&	0.229\%	&	0.263\%	&	0.195\%	&	0.161\%	\\
&	120	&	8.886	&	-0.810\%	&	0.158\%	&	0.180\%	&	0.124\%	&	0.090\%	\\
\hline														
$y=0.08$&	80	&	22.205	&	-12.069\%	&	6.071\%	&	2.891\%	&	0.270\%	&	-0.050\%	\\
&	90	&	16.207	&	-9.447\%	&	4.782\%	&	1.944\%	&	1.018\%	&	0.673\%	\\
&	100	&	11.704	&	-7.579\%	&	3.819\%	&	1.546\%	&	0.820\%	&	0.265\%	\\
&	110	&	8.367	&	-6.215\%	&	3.131\%	&	1.267\%	&	0.681\%	&	0.502\%	\\
&	120	&	5.93	&	-5.194\%	&	2.546\%	&	0.995\%	&	0.540\%	&	0.354\%	\\
\hline														
$y=0.04$&	80	&	20.35	&	-28.310\%	&	16.378\%	&	9.268\%	&	5.405\%	&	-0.010\%	\\
&	90	&	13.497	&	-23.494\%	&	19.434\%	&	2.052\%	&	0.563\%	&	2.430\%	\\
&	100	&	8.944	&	-19.857\%	&	16.167\%	&	2.102\%	&	1.409\%	&	-2.661\%	\\
&	110	&	5.912	&	-17.084\%	&	13.667\%	&	1.353\%	&	0.761\%	&	-0.507\%	\\
&	120	&	3.898	&	-14.956\%	&	11.852\%	&	1.411\%	&	0.770\%	&	-1.821\%	\\
\hline														
$y=0.00$&	80	&	20	&	-48.735\%	&	21.690\%	&	10.220\%	&	4.460\%	&	0.315\%	\\
&	90	&	11.697	&	-42.011\%	&	41.352\%	&	10.712\%	&	-1.470\%	&	2.240\%	\\
&	100	&	6.932	&	-36.440\%	&	38.344\%	&	-3.073\%	&	1.039\%	&	-3.390\%	\\
&	110	&	4.155	&	-31.986\%	&	33.069\%	&	-2.142\%	&	1.035\%	&	3.153\%	\\
&	120	&	2.51	&	-28.406\%	&	28.765\%	&	-2.112\%	&	-0.159\%	&	2.869\%	\\
\hline														
\end{tabular}	\\
error ratio =  100*(value-benchmark)/benchmark
\end{center}
\end{table}	
\clearpage	
										
\begin{table}[H]
\footnotesize
\begin{center}
\caption{American Calls ($T=3, K=100$)}\label{call3y}
%\vspace{.3cm}\\																			
\begin{tabular}{l|c|c|ccccc}																			
\hline
&	$S_0$	&	Benchmark	&	0th	&	1st	&	2nd	&	3rd	&	4th	\\
\hline
$\sigma=0.2$&	80	&	2.580	&	2.241	&	2.847	&	2.612	&	2.602	&	2.566	\\
$r=0.03$&	90	&	5.167	&	4.355	&	5.822	&	5.240	&	5.199	&	5.140	\\
$y=0.07$&	100	&	9.066	&	7.386	&	10.453	&	9.204	&	9.121	&	8.877	\\
&	110	&	14.443	&	11.331	&	17.036	&	14.763	&	14.566	&	14.322	\\
&	120	&	21.414	&	16.117	&	25.307	&	23.173	&	22.023	&	21.044	\\
\hline
$\sigma=0.4$&	80	&	11.326	&	10.309	&	11.998	&	11.475	&	11.399	&	11.373	\\
$r=0.03$&	90	&	15.722	&	14.162	&	16.769	&	15.975	&	15.858	&	15.770	\\
$y=0.07$&	100	&	20.793	&	18.532	&	22.318	&	21.132	&	20.948	&	20.851	\\
&	110	&	26.494	&	23.363	&	28.609	&	26.902	&	26.672	&	26.536	\\
&	120	&	32.781	&	28.598	&	35.599	&	33.390	&	33.000	&	32.681	\\
\hline
$\sigma=0.3$&	80	&	5.518	&	4.644	&	6.254	&	5.564	&	5.562	&	5.511	\\
$r=0.00$&	90	&	8.842	&	7.269	&	10.197	&	8.951	&	8.914	&	8.763	\\
$y=0.07$&	100	&	13.142	&	10.542	&	15.407	&	13.285	&	13.178	&	12.989	\\
&	110	&	18.453	&	14.430	&	22.004	&	18.655	&	18.726	&	18.144	\\
&	120	&	24.791	&	18.882	&	30.019	&	25.587	&	23.554	&	24.215	\\
\hline
$\sigma=0.3$&	80	&	12.146	&	12.133	&	12.148	&	12.148	&	12.147	&	12.146	\\
$r=0.07$&	90	&	17.368	&	17.343	&	17.373	&	17.374	&	17.372	&	17.372	\\
$y=0.03$&	100	&	23.348	&	23.301	&	23.359	&	23.360	&	23.357	&	23.355	\\
&	110	&	29.964	&	29.882	&	29.980	&	29.982	&	29.977	&	29.976	\\
&	120	&	37.104	&	36.972	&	37.130	&	37.134	&	37.125	&	37.120	\\
\hline
\end{tabular}\\
Number of simulations = 10,000,000, $\lambda=2$, Number of time steps = 6000
\vspace{15mm} ~ \\
Error ratio \\
\begin{tabular}{l|c|c|ccccc}														
\hline														
&	$S_0$	&	Benchmark	&	0th	&	1st	&	2nd	&	3rd	&	4th	\\
\hline														
$\sigma=0.2$&	80	&	2.58	&	-13.140\%	&	10.349\%	&	1.240\%	&	0.853\%	&	-0.543\%	\\
$r=0.03$&	90	&	5.167	&	-15.715\%	&	12.677\%	&	1.413\%	&	0.619\%	&	-0.523\%	\\
$y=0.07$&	100	&	9.066	&	-18.531\%	&	15.299\%	&	1.522\%	&	0.607\%	&	-2.085\%	\\
&	110	&	14.443	&	-21.547\%	&	17.953\%	&	2.216\%	&	0.852\%	&	-0.838\%	\\
&	120	&	21.414	&	-24.736\%	&	18.180\%	&	8.214\%	&	2.844\%	&	-1.728\%	\\
\hline														
$\sigma=0.4$&	80	&	11.326	&	-8.979\%	&	5.933\%	&	1.316\%	&	0.645\%	&	0.415\%	\\
$r=0.03$&	90	&	15.722	&	-9.922\%	&	6.659\%	&	1.609\%	&	0.865\%	&	0.305\%	\\
$y=0.07$&	100	&	20.793	&	-10.874\%	&	7.334\%	&	1.630\%	&	0.745\%	&	0.279\%	\\
&	110	&	26.494	&	-11.818\%	&	7.983\%	&	1.540\%	&	0.672\%	&	0.159\%	\\
&	120	&	32.781	&	-12.760\%	&	8.596\%	&	1.858\%	&	0.668\%	&	-0.305\%	\\
\hline														
$\sigma=0.3$&	80	&	5.518	&	-15.839\%	&	13.338\%	&	0.834\%	&	0.797\%	&	-0.127\%	\\
$r=0.00$&	90	&	8.842	&	-17.790\%	&	15.325\%	&	1.233\%	&	0.814\%	&	-0.893\%	\\
$y=0.07$&	100	&	13.142	&	-19.784\%	&	17.235\%	&	1.088\%	&	0.274\%	&	-1.164\%	\\
&	110	&	18.453	&	-21.801\%	&	19.243\%	&	1.095\%	&	1.479\%	&	-1.675\%	\\
&	120	&	24.791	&	-23.835\%	&	21.088\%	&	3.211\%	&	-4.990\%	&	-2.323\%	\\
\hline														
$\sigma=0.3$&	80	&	12.146	&	-0.107\%	&	0.016\%	&	0.016\%	&	0.008\%	&	0.000\%	\\
$r=0.07$&	90	&	17.368	&	-0.144\%	&	0.029\%	&	0.035\%	&	0.023\%	&	0.023\%	\\
$y=0.03$&	100	&	23.348	&	-0.201\%	&	0.047\%	&	0.051\%	&	0.039\%	&	0.030\%	\\
&	110	&	29.964	&	-0.274\%	&	0.053\%	&	0.060\%	&	0.043\%	&	0.040\%	\\
&	120	&	37.104	&	-0.356\%	&	0.070\%	&	0.081\%	&	0.057\%	&	0.043\%	\\
\hline														
\end{tabular}	\\							
error ratio =  100*(value-benchmark)/benchmark													
\end{center}
\end{table}
\clearpage

\begin{table}[H]
\footnotesize
\begin{center}
\caption{American Calls ($T=0.5, K=100, r=0.03, y=0.07$)}\label{call0.5y}
%\vspace{.3cm}\\																			
\begin{tabular}{l|c|c|ccccc}																			
\hline
&	$S_0$	&	Benchmark	&	0th	&	1st	&	2nd	&	3rd	&	4th	\\
\hline
$\sigma=0.2$&	80	&	0.219	&	0.215	&	0.222	&	0.220	&	0.220	&	0.220	\\
&	90	&	1.386	&	1.345	&	1.413	&	1.391	&	1.389	&	1.389	\\
&	100	&	4.783	&	4.578	&	4.920	&	4.807	&	4.795	&	4.791	\\
&	110	&	11.098	&	10.421	&	11.569	&	11.172	&	11.137	&	11.221	\\
&	120	&	20.000	&	18.302	&	20.536	&	20.350	&	20.238	&	20.092	\\
\hline
$\sigma=0.4$&	80	&	2.689	&	2.651	&	2.710	&	2.695	&	2.692	&	2.693	\\
&	90	&	5.722	&	5.622	&	5.778	&	5.736	&	5.729	&	5.730	\\
&	100	&	10.239	&	10.021	&	10.365	&	10.272	&	10.257	&	10.261	\\
&	110	&	16.181	&	15.768	&	16.424	&	16.241	&	16.214	&	16.232	\\
&	120	&	23.360	&	22.650	&	23.778	&	23.459	&	23.411	&	23.395	\\		
\hline
\end{tabular}\\
%\vspace{.3cm}
Number of simulations = 10,000,000, $\lambda=2$, Number of time steps = 1000
\vspace{1cm} ~ \\
Error ratio\\														
\begin{tabular}{l|c|c|ccccc}														
\hline														
&	$S_0$	&	Benchmark	&	0th	&	1st	&	2nd	&	3rd	&	4th	\\
\hline														
$\sigma=0.2$&	80	&	0.219	&	-1.826\%	&	1.370\%	&	0.457\%	&	0.457\%	&	0.457\%	\\
&	90	&	1.386	&	-2.958\%	&	1.948\%	&	0.361\%	&	0.216\%	&	0.216\%	\\
&	100	&	4.783	&	-4.286\%	&	2.864\%	&	0.502\%	&	0.251\%	&	0.167\%	\\
&	110	&	11.098	&	-6.100\%	&	4.244\%	&	0.667\%	&	0.351\%	&	1.108\%	\\
&	120	&	20	&	-8.490\%	&	2.680\%	&	1.750\%	&	1.190\%	&	0.460\%	\\
\hline														
$\sigma=0.4$&	80	&	2.689	&	-1.413\%	&	0.781\%	&	0.223\%	&	0.112\%	&	0.149\%	\\
&	90	&	5.722	&	-1.748\%	&	0.979\%	&	0.245\%	&	0.122\%	&	0.140\%	\\
&	100	&	10.239	&	-2.129\%	&	1.231\%	&	0.322\%	&	0.176\%	&	0.215\%	\\
&	110	&	16.181	&	-2.552\%	&	1.502\%	&	0.371\%	&	0.204\%	&	0.315\%	\\
&	120	&	23.36	&	-3.039\%	&	1.789\%	&	0.424\%	&	0.218\%	&	0.150\%   	\\
\hline														
\end{tabular}\\														
error ratio =  100*(value-benchmark)/benchmark													
\end{center}
\end{table}
\normalsize

As for the third order term, consulting the results based on IBP,
we are capable of determining an appropriate size of
the variance for each normal density applied in the approximation of a delta function.
Unfortunately, however,  this IBP method does not yield stable results for some cases in computing the fourth-order term.
It is clear that we want to use a small enough variance for the normal density so that it is a reasonable approximation 
of the delta function. On the other hand, too small variance increases the variation (dispersion) of simulation 
result. Therefore, we change the variance
from some large value to a smaller one gradually for a given number of simulation paths
and picks up the smallest value beyond which the variation (dispersion) starts to increase.

This scheme can be applied to general cases where the density functions of the underlying models
are not explicitly available. 
In fact, we adopt this approach for the numerical example for the Heston model in the next 
subsection~\footnote{
Notice that there is no need to use unnecessarily small variance for the approximation of delta function.
Intuitively speaking, the delta function within the expectation operation extracts the density where its
argument vanishes. Thus, as long as the density functions of the underlyings do not change significantly
within a given range, one can use it as a variance of the normal density function as an approximation of 
the corresponding delta function.}.
Of course, there is no guarantee that the choice of variance that gives the smallest dispersion in simulation 
also yields the smallest bias in the numerical result. 
However, the numerical result suggests that the method produces
accurate enough approximation for practical use given a reasonable number of simulation paths.
Note here that the large number of paths used in this example is to confirm the convergence of higher 
order expansions. For practical pricing purpose, as can be seen in the following example of Heston model, 
there is no need to run such a large number of simulation trials.

Table \ref{put3y} presents the result  for American put options
with $T=3$, $K=100$, $\sigma=0.2$ and $r=0.08$, which 
confirms that the error ratios 
become improved in the results up to the third or the fourth order comparing with those
up to the first and the second orders.
In total, the approximations up to the fourth order provide the most precise ones
in terms of the error ratios.
Note also that for the dividend rate $y=0.12$ and $0.07$,
adding the fourth order term to the third one
makes the accuracies of the approximations improved,
while for $y=0.04$ and $0.00$, it makes the accuracies worse in
three and four out of the five cases, respectively.
Table \ref{call3y} ($T=3, K=100$) and Table \ref{call0.5y} ($T=0.5, K=100, r=0.03, y=0.07$) present the results  for American call options,
which shows 
the error ratios become 
smaller by adding the third or/and fourth order terms.

%%%%%%%%%%%%%%%%%%%%%%%%%%%%%%%%%%%%%%%%%%%%%%%%%%%%%%%%
\subsection{Example 2: Heston model} 
%%%%%%%%%%%%%%%%%%%%%%%%%%%%%%%%%%%%%%%%%%%%%%%%%%%%%%%%%
The next example takes Heston model (\ref{Heston}):
\begin{eqnarray}\label{Heston}
d S_t &=& (r-y) S_t dt + S_t \sqrt{\nu_t} \cdot  d W_{1,t} \\
d \nu_t &=& \xi (\theta- \nu_t) dt + \eta\sqrt{\nu_t} (\rho d W_{1,t} + \sqrt{1-\rho^2} d W_{2,t})~;\quad \nu_0 = \sigma^2,
\end{eqnarray}
where $\xi$, $\eta$ and $\theta$ are positive constants such that $\xi\eta \geq \theta^2/2$ and $W_{1,t} \bot W_{2,t}$.
Then, we compute the approximate values for American put prices
($T=0.25,~0.5$, $K=100$, $r=0.05$, $y=0$, $\eta=0.1$, $\xi=3.0$, $\theta=0.04$) up to the third order
based on our scheme, (\ref{V1st}), (\ref{V2nd}) and (\ref{V3rd}) with 50,000 trials in Monte Carlo simulation.
Here, we adopt \cite{BelN} as the benchmark values,
in which a two-dimensional tree with two hundred time steps 
and a Control Variate technique is applied.
Moreover, an asymptotic expansion method, particularly,
the equation appearing in p.113 of
\cite{asymptotic4} is used for computation of European option prices,
 that is, 0-th order $v^{(0)}$.

Table \ref{heston_put} demonstrates that our method works effectively in the Heston model,
which suggests its applicability to the pricing problem of American options for other multi-dimensional models, too.
The numerical result shows that the expansion up to the third order improves the accuracies in most of the cases. 
For the choice of the normal density as a approximate delta function, we have used the variance 
found to work well in the previous BS model.
In the case where it produces too much dispersion in simulation, we have applied the general methodology
explained in the previous subsection to pick up an appropriate variance.
The example shows that the relatively small number of simulation trials is enough to obtain a reasonable 
accuracy for the practical use.
In Table~\ref{heston_put_largepath}, we have given the numerical results with larger number of simulations $500,000$
for the same set of American options with $T=0.5$ in Table~\ref{heston_put}. Although the
improvement of accuracy from the second to the third order approximation becomes more robust in this case,
one can observe that the size of the change in option prices is rather small. 
\\
\\
{\bf{Remark}:}\\ Although higher order integration is required, the direct evaluation of 
(\ref{V2ndDirect}) ( and corresponding expressions in other orders ) is also possible once we 
know the transition density of the underlying states. For the diffusion models, 
it is always possible to obtain approximation using the asymptotic expansion~\cite{T}.
If there exists efficient enough integration technique, such as Gaussian quadrature and its extension, it could provide 
another pricing technique. In fact, in BS model, we have compared this semi-analytic
results (by brute force integration within $\pm 5$-sigma range and using a normal density function with variance of 
$1$bp of the stock process at each time as an approximation for the delta function) 
to those obtained from the particle method up to the second order terms. We confirmed the consistency
between their numerical results.

\begin{table}[H]
\scriptsize
\begin{center}
\caption{American Puts in Heston model ($K=100, r=0.05, y=0, \eta=0.1, \xi=3.0, \theta=0.04$)}\label{heston_put}
%\vspace{.3cm}\\																			
$T=0.25$ \\
\begin{tabular}{l|c|c|cccc|cccc}																			
\hline
&	$S_0$	& Benchmark	&	0th	&	1st	&	2nd	&	3rd & ER(0th) & ER(1st) & ER(2nd) & ER(3rd) \\
\hline
$\rho=-0.1$	&	90	&	10.171	&	9.643	&	10.507	&	10.293	&	10.141	&	-5.18	\% &	3.31	\% &	1.20	\% &	-0.29	\% \\
$\sigma=0.2$	&	100	&	3.475	&	3.374	&	3.556	&	3.486	&	3.481	&	-2.89	\% &	2.35	\% &	0.34	\% &	0.19	\% \\
	&	110	&	0.774	&	0.758	&	0.783	&	0.775	&	0.775	&	-2.03	\% &	1.21	\% &	0.18	\% &	0.14	\% \\
\hline
$\rho=-0.7$	&	90	&	10.121	&	9.573	&	10.455	&	10.253	&	10.101	&	-5.41	\% &	3.31	\% &	1.31	\% &	-0.19	\% \\
$\sigma=0.2$	&	100	&	3.481	&	3.383	&	3.559	&	3.493	&	3.487	&	-2.81	\% &	2.24	\% &	0.36	\% &	0.18	\% \\
	&	110	&	0.842	&	0.829	&	0.854	&	0.845	&	0.845	&	-1.53	\% &	1.43	\% &	0.37	\% &	0.41	\% \\
\hline
$\rho=-0.1$	&	90	&	12.182	&	11.896	&	12.347	&	12.190	&	12.173	&	-2.35	\% &	1.36	\% &	0.07	\% &	-0.08	\% \\
$\sigma=0.4$	&	100	&	6.496	&	6.379	&	6.572	&	6.504	&	6.501	&	-1.80	\% &	1.17	\% &	0.12	\% &	0.08	\% \\
	&	110	&	3.091	&	3.047	&	3.118	&	3.092	&	3.092	&	-1.43	\% &	0.85	\% &	0.03	\% &	0.02	\% \\
\hline
$\rho=-0.7$	&	90	&	12.112	&	11.832	&	12.291	&	12.132	&	12.116	&	-2.31	\% &	1.47	\% &	0.16	\% &	0.03	\% \\
$\sigma=0.4$	&	100	&	6.490	&	6.377	&	6.565	&	6.505	&	6.503	&	-1.74	\% &	1.16	\% &	0.23	\% &	0.20	\% \\
	&	110	&	3.146	&	3.104	&	3.180	&	3.157	&	3.157	&	-1.31	\% &	1.11	\% &	0.36	\% &	0.37	\% \\
\hline
\end{tabular}\\
Number of simulations = 50,000, $\lambda=4$, Number of time steps = 2000\\
ER =  100*(value-benchmark)/benchmark
~ \\
~ \\
$T=0.5$ \\
\begin{tabular}{l|c|c|cccc|cccc}																			
\hline
&	$S_0$	& Benchmark	&	0th	&	1st	&	2nd	&	3rd & ER(0th) & ER(1st) & ER(2nd) & ER(3rd) \\
\hline
$\rho=-0.1$	&	90	&	10.648	&	9.864	&	11.236	&	10.752	&	10.532	&	-7.36	\% &	5.53	\% &	0.98	\% &	1.09	\% \\
$\sigma=0.2$	&	100	&	4.647	&	4.423	&	4.835	&	4.676	&	4.665	&	-4.83	\% &	4.04	\% &	0.61	\% &	0.38	\% \\
	&	110	&	1.683	&	1.624	&	1.733	&	1.692	&	1.693	&	-3.50	\% &	2.94	\% &	0.55	\% &	0.55	\% \\
\hline
$\rho=-0.7$	&	90	&	10.564	&	9.766	&	11.183	&	10.688	&	10.490	&	-7.55	\% &	5.87	\% &	1.18	\% &	0.70	\% \\
$\sigma=0.2$	&	100	&	4.664	&	4.443	&	4.844	&	4.684	&	4.678	&	-4.73	\% &	3.88	\% &	0.43	\% &	0.31	\% \\
	&	110	&	1.787	&	1.732	&	1.837	&	1.798	&	1.797	&	-3.08	\% &	2.79	\% &	0.61	\% &	0.52	\% \\
\hline
$\rho=-0.1$	&	90	&	13.314	&	12.712	&	13.664	&	13.375	&	13.283	&	-4.52	\% &	2.63	\% &	0.46	\% &	0.23	\% \\
$\sigma=0.4$	&	100	&	8.008	&	7.705	&	8.207	&	8.070	&	8.021	&	-3.78	\% &	2.48	\% &	0.77	\% &	0.16	\% \\
	&	110	&	4.545	&	4.399	&	4.642	&	4.567	&	4.550	&	-3.21	\% &	2.12	\% &	0.48	\% &	0.09	\% \\
\hline
$\rho=-0.7$	&	90	&	13.217	&	12.625	&	13.602	&	13.314	&	13.229	&	-4.48	\% &	2.91	\% &	0.73	\% &	0.09	\% \\
$\sigma=0.4$	&	100	&	8.000	&	7.705	&	8.196	&	8.048	&	8.012	&	-3.69	\% &	2.46	\% &	0.60	\% &	0.15	\% \\
	&	110	&	4.620	&	4.479	&	4.709	&	4.627	&	4.612	&	-3.04	\% &	1.93	\% &	0.16	\% &	0.17	\% \\
\hline
\end{tabular}\\
Number of simulations = 50,000, $\lambda=8$, Number of time steps = 2000\\
ER =  100*(value-benchmark)/benchmark													
%\vspace{.3cm}
\end{center}
\end{table}
\normalsize

\begin{table}[!htp]
\scriptsize
\begin{center}
\caption{The same setup with  $T=0.5$ in Table~\ref{heston_put} but using larger number of simulation.}
\label{heston_put_largepath}
$T=0.5$ \\
\begin{tabular}{l|c|c|cccc|cccc}																			
\hline
&	$S_0$	& Benchmark	&	0th	&	1st	&	2nd	&	3rd & ER(0th) & ER(1st) & ER(2nd) & ER(3rd) \\
\hline
$\rho=-0.1$	&	90	&	10.648	&	9.864	&	11.259	&	10.758	&	10.540	&	-7.36	\% &	5.74	\% &	1.03	\% &	-1.01	\% \\
$\sigma=0.2$	&	100	&	4.647	&	4.423	&	4.831	&	4.674	&	4.653	&	-4.83	\% &	3.95	\% &	0.57	\% &	0.11	\% \\
	&	110	&	1.683	&	1.624	&	1.729	&	1.688	&	1.683	&	-3.50	\% &	2.70	\% &	0.28	\% &	-0.02	\% \\
\hline
$\rho=-0.7$	&	90	&	10.564	&	9.766	&	11.173	&	10.696	&	10.457	&	-7.55	\% &	5.77	\% &	1.26	\% &	-1.01	\% \\
$\sigma=0.2$	&	100	&	4.664	&	4.443	&	4.841	&	4.688	&	4.672	&	-4.73	\% &	3.81	\% &	0.53	\% &	0.19	\% \\
	&	110	&	1.787	&	1.732	&	1.839	&	1.800	&	1.795	&	-3.08	\% &	2.86	\% &	0.71	\% &	0.44	\% \\
\hline
$\rho=-0.1$	&	90	&	13.314	&	12.712	&	13.676	&	13.384	&	13.311	&	-4.52	\% &	2.72	\% &	0.53	\% &	-0.02	\% \\
$\sigma=0.4$	&	100	&	8.008	&	7.705	&	8.202	&	8.043	&	8.007	&	-3.78	\% &	2.41	\% &	0.44	\% &	-0.01	\% \\
	&	110	&	4.545	&	4.399	&	4.643	&	4.561	&	4.546	&	-3.21	\% &	2.14	\% &	0.34	\% &	0.02	\% \\
\hline
$\rho=-0.7$	&	90	&	13.217	&	12.625	&	13.582	&	13.292	&	13.216	&	-4.48	\% &	2.76	\% &	0.57	\% &	-0.01	\% \\
$\sigma=0.4$	&	100	&	8.000	&	7.705	&	8.194	&	8.039	&	8.003	&	-3.69	\% &	2.42	\% &	0.49	\% &	0.04	\% \\
	&	110	&	4.620	&	4.479	&	4.718	&	4.640	&	4.625	&	-3.04	\% &	2.12	\% &	0.43	\% &	0.10	\% \\
\hline
\end{tabular}\\
Number of simulations = 500,000, $\lambda=8$, Number of time steps = 2000\\
ER =  100*(value-benchmark)/benchmark													
%\vspace{.3cm}
\end{center}
\end{table}				
\normalsize

%%%%%%%%%%%%%%%%%%%%%%%%%%%%%%%%%%%%%%%%%%%%%%%%%%%%%%%%
\section{Conclusions}
%%%%%%%%%%%%%%%%%%%%%%%%%%%%%%%%%%%%%%%%%%%%%%%%%%%%%%%%
This paper proposed a new calculation technique for American options in an FBSDE framework.
The well-known decomposition of an American option price with that of the corresponding 
European option and additional early exercise premium can be written in a form of 
a decoupled non-linear FBSDE. We have used the recently proposed perturbation technique 
of FBSDE with an interacting particle method to obtain numerical results.
We have tested the effectiveness of our approximation by comparing the numerical results to those
obtained from existing tree algorithms.
Although there remains some subtlety for choosing an appropriate variance for the normal density function 
as a proxy of the Dirac delta function, the proposed method for the variance choice yields accurate enough approximations
 for BS as well as Heston models.
In the paper, we could only test a narrow range of parameters
with relatively short expiries of options due to the limitation of existing benchmark results. However, the results are quite encouraging to suggest that our perturbation technique combined with an interacting particle method can be applied to much broader range of models and parameters.

%%%%%%%%%%%%%%%%%%%%%%%%%%%%%%%%%%%%%%%%%%%%%%%%%%%%%%%%%%%%%%%%%%%%%%%%%%%%%

%%%%%%%%%%%%%%%%%%%%%%%%%%%%%%%%%%%%%%%%%%%%%%%%%%%%%%%%%%%%%%%%%%%%

\end{document}